\documentclass[12pt]{article}
\usepackage{amsfonts}
\usepackage{amssymb}
\usepackage{amsbsy}
\usepackage{mathrsfs}
\usepackage{amsmath}
\usepackage{amsthm}
\usepackage{graphicx}    
\usepackage{rotating}    
\usepackage{epsfig}
\usepackage{color}       

\input epsf.sty

\newlength{\TZ}
\newcommand{\BEQ}{\begin{equation}}     
\newcommand{\BEA}{\begin{eqnarray}}
\newcommand{\BD}{\begin{displaymath}}
\newcommand{\EEQ}{\end{equation}}       
\newcommand{\EEA}{\end{eqnarray}}
\newcommand{\ED}{\end{displaymath}}
\newcommand{\D}{{\rm d}}                
\newcommand{\II}{{\rm i}}               
\newcommand{\demi}{\frac{1}{2}}         
\newcommand{\wit}[1]{\widetilde{#1}}    
\newcommand{\wht}[1]{\widehat{#1}}      
\newcommand{\lap}[1]{\overline{#1}}     

\renewcommand{\vec}[1]{\boldsymbol{#1}} 

\newcommand{\appsektion}[1]{\setcounter{equation}{0}\setcounter{subsection}{0}\setcounter{table}{0}\setcounter{figure}{0}
\section*{Appendix. #1}
\renewcommand{\theequation}{A.\arabic{equation}}
              \renewcommand{\thesection}{A}\renewcommand{\thetable}{A\arabic{table}}\renewcommand{\thefigure}{A.\arabic{figure}} }

\newcommand{\R}{\mathbb{R}}

\catcode`\@=11
\def\numberbysection{\@addtoreset{equation}{section}
        \def\theequation{\thesection.\arabic{equation}}}
\numberbysection

\begin{document}

\begin{titlepage}

\vskip 1.5 cm
\begin{center}
{\LARGE \bf Dynamical symmetries in the non-equilibrium dynamics of the directed spherical model}
\end{center}

\vskip 2.0 cm
\centerline{{\bf Malte Henkel}$^{a,b}$ and {\bf Stoimen Stoimenov}$^c$}
\vskip 0.5 cm
\centerline{$^a$Laboratoire de Physique et Chimie Th\'eoriques (CNRS UMR 7019),}
\centerline{Universit\'e de Lorraine Nancy, B.P. 70239, F -- 54506 Vand{\oe}uvre l\`es Nancy Cedex, France}
\vspace{0.5cm}
\centerline{$^b$Centro de F\'{i}sica Te\'{o}rica e Computacional, Universidade de Lisboa,}
\centerline{Campo Grande, P--1749-016 Lisboa, Portugal}
\vspace{0.5cm}
\centerline{$^c$ Institute of Nuclear Research and Nuclear Energy, Bulgarian Academy of Sciences,}
\centerline{72 Tsarigradsko chaussee, Blvd., BG -- 1784 Sofia, Bulgaria}
\vspace{0.5cm}

\begin{abstract}
The dynamical scaling and ageing in the relaxational dynamics of the quenched directed spherical model is analysed. 
The exact two-time correlation and response functions display new regimes of ballistic or anisotropic ballistic scaling, 
at larger distances than probed in the usual regime of diffusive scaling.
The r\^ole of long-ranged initial correlations on the existence of these scaling regimes is clarified.
Their dynamical symmetries are described in terms of extensions of the Schr\"odinger algebra appropriate to non-equilibrium dynamics  
in that the anisotropic ballistic scaling regime can be interpreted in terms of meta-Schr\"odinger invariance
while the regime of isotropic ballistic scaling is meta-conformally invariant.
\end{abstract}
\end{titlepage}

\setcounter{footnote}{0}

\section{Introduction} \label{sec:intro}

Physical ageing in a many-body system can arise after a quenching it from some initial (disordered) state
either onto a critical point or into a two-phase coexistence regime
\cite{Struik78,Bray94a,Cugl03,Henkel10,Taeu14}. In such studies, dynamical symmetries play an important r\^ole.
A central quantity is the time-dependent length scale $L=L(t)$, which for large times typically
grows algebraically according to $L(t)\sim t^{1/z}$ where $z$
is the dynamical exponent.\footnote{Glassy systems with a logarithmic growth $L(t)\sim \ln^{\psi} t$ \cite{Cugl03} are not considered in this paper.}
Hence for large times, ageing systems are naturally driven to a state
which satisfies dynamical scaling. Relevant physical observables include two-time correlators $C$ and and two-time responses $R$.
In the limit of large times $t,s\gg \tau_{\rm micro}$ (with a microscopic reference time $\tau_{\rm micro}$) such that $t/s>1$,
they are expected to admit dynamical scaling forms
\begin{subequations} \label{1.1}
\begin{align} \label{1.1a}
C(t,s;\vec{r}) &:= \left\langle \phi(t,\vec{r}) \phi(s,\vec{0}) \right\rangle = s^{-b} F_C\left(\frac{t}{s},\frac{|\vec{r}|^z}{t+s}\right) \\
R(t,s;\vec{r}) &:= \left.\frac{\delta\bigl\langle \phi(t,\vec{r})\bigr\rangle}{\delta h(s,\vec{0})}\right|_{h=0}
                 = s^{-1-a} F_R\left(\frac{t}{s},\frac{|\vec{r}|^z}{t-s}\right) \label{1.1b}
\end{align}
\end{subequations}
where $\phi(t,\vec{r})$ is the (coarse-grained) time-space-dependent order parameter,
$h(s,\vec{r})$ is a sym\-me\-try-break\-ing external field and $a,b$ are system-dependent, but universal, ageing exponents.
Dynamical scaling by itself does not provide any information on the {\em form} of these universal scaling functions
(in (\ref{1.1}), spatial translation-invariance was tacitly admitted),
but in many cases one finds an asymptotic algebraic behaviour of the form
\BEQ \label{1.2}
F_C(y,0) \sim y^{-\lambda_C/z} \;\; , \;\; F_R(y,0) \sim y^{-\lambda_R/z}
\EEQ
for $y=t/s\gg 1$ and where $\lambda_C$ and $\lambda_R$ are the (universal) auto-correlation and auto-response exponents, respectively.
Further dynamical symmetries beyond scaling are required in order to fix the form of scaling functions $F_C$ and $F_R$ in (\ref{1.1}).

\renewcommand{\arraystretch}{1.2}
\begin{sidewaystable}
\begin{tabular}{|l|lll|l|c|l|} \hline
group                             & \multicolumn{3}{l|}{coordinate transformations}                                & co-variance & $\mathscr{S}$ & {\small abbreviations} \\
\hline
{\small ortho-conformal $(1+1)D$} & $z'=f(z)$    & $\bar{z}'=\bar{z}$                        &
              & correlator  & $4\partial_z \partial_{\bar{z}}=\partial_t^2+\partial_r^2$ & {\small$z=t+\II r$} \\
                                  & $z'=z$       & $\bar{z}'=\bar{f}(\bar{z})$               &                     & & & {\small$\bar{z}=t-\II r$} \\ \hline
{\small conformal galilean}       & $t'=b(t)$    & \multicolumn{2}{l|}{$\vec{r}'=\vec{r}\:\dot{b}(t)$}             & & & \\
                                  & $t'=t$       & $\vec{r}'=\vec{r}+\vec{a}(t)$             &                     & correlator & & \\
                                  & $t'=t$       & $\vec{r}'=\mathscr{R}(t)\vec{r}$          &                     & & & \\ \hline
{\small meta-conformal $1D$}      & $u=b(u)$     & $v'=v$                                    &
              & correlator & $\partial_t - \frac{1}{\beta}\partial_{r_{\|}}$ & {\small$u=t$} \\
                                  & $u'=u$       & $v'=c(v)$                                 &                     & & & {\small $v=t+\beta r_{\|}$} \\ \hline
{\small meta-conformal $2D$}      & $\tau'=\tau$ & $w'=f(w)$                                 & $\bar{w}'=\bar{w}$  & & & {\small$\tau=t$} \\
                                  & $\tau'=\tau$ & $w'=w$                                    & $\bar{w}'=\bar{f}(\bar{w})$
              & correlator & $\partial_t - \frac{1}{\beta}\partial_{r_{\|}}$ & {\small$w=t+\beta\bigl(r_{\|}+\II r_{\perp}\bigr)$} \\
                                  & $\tau'=b(\tau)$ & $w'=w$                                 & $\bar{w}'=\bar{w}$
              & & & {\small$\bar{w}=t+\beta\bigl(r_{\|}-\II r_{\perp}\bigr)$}\\[0.12cm] \hline
{\small Schr\"odinger-Virasoro}   & $t'=b(t)$    & \multicolumn{2}{l|}{$\vec{r}'=\vec{r}\,\sqrt{\dot{b}(t)\,}$}    & & & \\
                                  & $t'=t$       & $\vec{r}'=\vec{r}+\vec{a}(t)$             &
              & response   & $\partial_t - \frac{1}{2{\cal M}}\Delta_{\vec{r}}$ & \\
                                  & $t'=t$       & $\vec{r}'=\mathscr{R}(t)\vec{r}$          &                     & & & \\[0.12cm] \hline
{\small meta-Schr\"odinger-Virasoro} & $t'=b(t)$ & $v'=v$ & \multicolumn{1}{l|}{$\vec{r}_{\perp}'=\vec{r}_{\perp}\,\sqrt{\dot{b}(t)\,}$} & & & \\
                                  & $t'=t$       & $v'=c(v)$                                 & $\vec{r}_{\perp}'=\vec{r}_{\perp}$
              & response   & $\partial_t - \frac{1}{\beta}\partial_{r_{\|}}- \frac{1}{2{\cal M}}\Delta_{\vec{r}_{\perp}}$ & {\small$v=t+\beta r_{\|}$} \\
                                  & $t'=t$       & $r_{\|}'=r_{\|}$                          & $\vec{r}_{\perp}'=\vec{r}_{\perp}+\vec{a}(t)$    & & & \\
                                  & $t'=t$       & $r_{\|}'=r_{\|}$                          & $\vec{r}_{\perp}'=\mathscr{R}(t)\vec{r}_{\perp}$ & & & \\ \hline
\end{tabular}
\caption[tab1]{{\small
Examples of infinite-dimensional groups of time-space transformations, defined through abstract coordinate t
transformations on time ($t$) and space
coordinates ($r\in\mathbb{R}$ or $\vec{r}\in\mathbb{R}^d$).
A physical bias is parametrised by the constant $\beta\ne 0$ and distinguishes a preferred spatial direction
$r_{\|}\in\mathbb{R}$ from transverse spatial directions $\vec{r}_{\perp}\in\mathbb{R}^{d-1}$.
The expressions of the time-space transformations can be simplified through
several abbreviations, wherein $f,\bar{f},b,c,\vec{a}$
are differentiable (vector-valued) functions of their argument, $\dot{b}(t)=\D b(t)/\D t$
and $\mathscr{R}(t)\in\mbox{\sl SO}(d)$ is a time-dependent rotation matrix.
The physical nature of covariant $n$-point functions as correlators or responses is indicated.
$\mathscr{S}$ is the invariant Schr\"odinger operator, where $\Delta_{\vec{r}}$ is
the spatial laplacian.}
\label{tab1}}
\end{sidewaystable}

For equilibrium phase transitions, scale-invariance can in many situations be generalised to conformal invariance, which especially for two-dimensional
systems leads to spectacular results \cite{Belavin84,Francesco97,Henkel12,Rychkov17}.
For non-equilibrium systems undergoing ageing, dynamical scaling holds true \cite{Struik78}.
In particular, in the case of phase-ordering kinetics, after a quench to $T<T_c$ and a microscopic dynamics without any conservation laws (`model A'), the dynamical
exponent $z=2$ \cite{Bray94a,Bray94b,Bray94c}.
Then Schr\"odinger-invariance is a good candidate for an extended set of local time-space-dependent scale transformations \cite{Henkel94}, 
in the regime of large times $t$ and large spatial distances $|\vec{r}|$ such that $|\vec{r}|^2/t$ is kept fixed.
The Schr\"odinger Lie algebra \cite{Jacobi1842,Lie1881}
is the maximal finite-dimensional Lie sub-algebra of the infinite-dimensional Schr\"odinger-Virasoro Lie algebra \cite{Henkel03a,Unterberger12}.
See \cite{Henkel10,Henkel17c} for tutorials and detailed reviews on the applications and tests of Schr\"odinger-invariance 
on non-equilibrium relaxation phenomena and ageing.
Schr\"odinger-invariance furnishes the `best known description' of recent
experiments on the phase-ordering kinetics in liquid crystals \cite{Almeida21}.

Do there exist other possibilities than Schr\"odinger-invariance in scale-invariant non-equi\-li\-bri\-um dynamics~? 
Which physical conditions should be imposed~? 
Indeed, many known examples exist for non-trivial dynamical symmetries, see \cite{Duval09}.
Table~\ref{tab1} gives a list of infinite-dimensional time-space transformation groups which include dynamical scaling, with either $z=1$
(for ortho- \cite{Belavin84} and meta-conformal \cite{Henkel19a} or conformal galilean groups \cite{Havas78,Barnich07,Bagchi10})
or else $z=2$ (for the Schr\"odinger-related groups). Using dualisation techniques borrowed from
string theory, the physical nature of the co-variant $n$-point functions as either correlators or response functions can be derived
\cite{Henkel03a,Henkel14a,Henkel15,Henkel16,Henkel20}, as indicated in table~\ref{tab1}.
Here, we are interested in those transformations, called `meta-conformal' or `meta-Schr\"odinger', whose invariant Schr\"odinger operator
$\mathscr{S}$ admits a biased ballistic transport in a preferred direction (when $\beta\ne 0$).
Such biases may arise in the dynamics of spin systems with directed dynamics \cite{Schmittmann95,Marro99}. 
Examples might be massive particles in  a gravitational field or electrically charged particles in an external electric field 
(used to model fast ionic conductors) \cite{Katz83,Katz84,Chandra81}. Their non-equilibrium directed dynamics
has been studied intensively \cite{Dutta11}, notably in the directed Ising \cite{Godreche11,Godreche15a,Godreche15b,Godreche18} and the directed spherical
models \cite{Godreche13}. From the exact solutions of the $1D$ directed Glauber-Ising model
\cite{Godreche11,Godreche15a} and the directed spherical model in $d>2$ dimensions
\cite{Godreche13}, the dynamics of the single-time correlator $C(t,t;\vec{r})$ is found to be independent of the bias, whereas in the $2D$ directed Glauber-Ising model
quenched to $T=0$ a new dynamic behaviour is found \cite{Godreche15b,Godreche18}. Our main interest here is to see whether these models can be viewed as examples
of dynamical non-equilibrium symmetries with a bias. 
The required abstract constructions of those Lie groups have been carried out by us recently \cite{Stoimenov22}. In particular, one must
consider in detail the different time-space scaling required for the preferred and transverse directions:
\begin{enumerate}
\item for a \underline{meta-conformal symmetry}, one has an isotropic and ballistic scaling in both the preferred and the transverse directions,
such that both temporal coordinates $t$ and spatial coordinates $r_{\|}$ and $\vec{r}_{\perp}$ become large such that
\BEQ \label{gl:metaC}
\bigl| r_{\|}\bigr|/t = \mbox{\rm\small cste.} \;\; , \;\; \bigl| \vec{r}_{\perp}\bigr|/t = \mbox{\rm\small cste.}
\EEQ
which means that the dynamical exponent $z=1$ isotropically in all directions.
\item for a \underline{meta-Schr\"odinger symmetry}, the scaling in the preferred and transverse directions is different and one must take the scaling limit such that
\BEQ \label{gl:metaSchr}
\bigl| r_{\|}\bigr|/t = \mbox{\rm\small cste.} \;\; , \;\; \bigl| \vec{r}_{\perp}\bigr|^2 /t = \mbox{\rm\small cste.}
\EEQ
and consequently, there is an anisotropic dynamical exponent $z_{\|}=1$ in the preferred direction whereas $z_{\perp}=2$ in the transverse directions.
\end{enumerate}
Clearly, these scaling limits are different from the usual one, namely $|r_{\|}|^2 /t\sim |\vec{r}_{\perp}|^2 /t=\mbox{\rm\small cste.}$ 
required for the Schr\"odinger-invariant systems of standard phase-ordering kinetics. 
Is it possible to find evidence for the existence of either scaling limit (\ref{gl:metaC},\ref{gl:metaSchr}),
when considering much larger distances than those examined in studies of Schr\"odinger-invariance~?

Indeed, in the directed $1D$ Glauber-Ising model, quenched to $T=0$ from a disordered initial state, the scaling of the two-time correlator is consistent with
meta-conformal invariance and ballistic scaling with $z=1$, provided that the initial state admits long-range correlations
$C(0,r)\sim \bigl|r\bigr|^{-\aleph}$ with $0<\aleph<\demi$ \cite{Henkel19a}.
For $\aleph>\demi$, the scaling behaviour reverts to the habitual one, described by Schr\"odinger-invariance and $z=2$. In order to see if long-range
initial correlations are an important ingredient for ballistic scaling behaviour, we shall analyse the directed spherical model, again with long-ranged
initial correlations in the preferred direction in order to better appreciate the influence of the transverse directions and the respective scaling limits
(\ref{gl:metaC}) and (\ref{gl:metaSchr}), respectively. Since the spherical model is solved exactly in Fourier space and the back-transformation to direct space
involves tickly questions of how to take the appropriate scaling limit, we prefer to carry out these tests in Fourier space.
We shall find the conditions such that the biased spherical model in $d>2$ spatial dimensions becomes only the second example of a physical realisation of
meta-conformal invariance or, in different parts of parameter space, the first physical realisation of meta-Schr\"odinger invariance. 

In these tests, an important technical point has to be made. The algebras in table~\ref{tab1} are obtained from their defining `standard' representation
of time-space transformations. In these representations, time-translations have the generator $X_{-1}=-\partial_t$. However, the specific form of the
order-parameter equation of motion in the biased spherical model 
$\bigl( \partial_t -\frac{1}{\beta}\partial_{\|}-\frac{1}{2{\cal M}}\Delta_{\vec{r}_{\perp}} - \mathfrak{z}(t)\bigr)\phi = \eta$ 
requires a different representation. Herein, the Lagrange multiplier $\mathfrak{z}(t)$ must be found from the spherical constraint. 
Asymptotically, for $t\to\infty$, this yields $\mathfrak{z}(t) \simeq {\digamma}/{(2t)}$ (higher-order terms would produce corrections to scaling). 
Hence, one must go over to a new `ageing' representation, by mapping the generators via $X_n \mapsto \lap{X}_n = e^{\xi \ln t} X_n e^{-\xi \ln t}$ etc. \cite{Stoimenov22}. 
The breaking of time-translation-invariance in non-equilibrium dynamics is now described through the generator $\lap{X}_{-1} = -\partial_t + \xi/t$. 
Furthermore, the transformed Schr\"odinger operator $\mathscr{S} \mapsto \lap{\mathscr{S}} = e^{\xi \ln t} \mathscr{S} e^{-\xi \ln t} = \mathscr{S} - \xi/t$,
where $\mathscr{S}$ is taken from table~\ref{tab1}, now contains the required $1/t$-potential.

This work contains a statistical mechanics part and a dynamical symmetry part, and is organised as follows. 
In the statistical mechanics part, section~\ref{sec:spherique} recalls 
the formulation and exact solution of the directed spherical model  before we give in section~\ref{sec:reponse-corr} its 
exact two-time correlators and responses, in the presence of long-range initial correlations.
In the dynamical symmetry part, 
section~\ref{sec:generators} lists the meta-conformal and meta-Schr\"odinger generators in Fourier space, 
both for the standard representation as well as for the `ageing' representation, the latter one being required for applications to non-equilibrium dynamics. 
Then section~\ref{sec:cov-2point} gives the predictions for the co-variant two-point functions.  
Finally, the results of both parts are compared in section~\ref{sec:comparaison} 
where we identify the conditions for realising either meta-conformal or meta-Schr\"odinger invariance in the spherical model.
We conclude in section~\ref{sec:concludere}.
An appendix gives the details of the asymptotic calculations in the directed spherical model.

\section{Spherical model} \label{sec:spherique}

The spherical model with a directional bias will serve as a case study on systems with meta-conformal or meta-Schr\"odinger invariance.
To make this paper more self-contained, we begin by briefly re-tracing the main steps in the derivation of two-time correlators and
responses, following \cite{Godreche13}.
The model is defined in terms of a continuous spin variable $S(t,\vec{n})\in\mathbb{R}$, subject to the (mean) spherical constraint
\BEQ
\sum_{\vec{n}\in\Lambda} \left\langle S(t,\vec{n})^2 \right\rangle = {\cal N}
\EEQ
where ${\cal N}=|\Lambda|$ is the number of sites of the hyper-cubic lattice $\Lambda\subset \mathbb{Z}^d$. Taking directly the
the infinite-size limit ${\cal N}\to\infty$, the spin variable becomes $S=S(t,\vec{r})\in\mathbb{R}$ which obey the defining equations of motion
(with the habitual scalings)
\BEQ \label{Langeq}
\partial_t S(t,\vec{r}) = \sum_{a=1}^d \biggl[ (1+v_a) S(t,\vec{r}-\vec{e}_a) + (1-v_a) S(t,\vec{r}+\vec{e}_a) - 2 S(t,\vec{r}) \biggr]
+ \mathfrak{z}(t) S(t,\vec{r}) + \eta(t,\vec{r})
\EEQ
where $\eta=\eta(t,\vec{r})$ is the usual centred gaussian white noise and $\mathfrak{z}(t)$ is the Lagrange multiplier which is to be found from the
spherical constraint. Furthermore, the directional bias is described by $\vec{v}$ and $\vec{e}_a$ is the unit vector in direction $a=1,\ldots,d$.

The formal solution is an immediate generalisation of the non-biased case \cite{Godr00b,Picone02}. In Fourier space, one has
\BEQ\label{solcor}
\wht{S}(t,\vec{q}) = e^{-\Omega(\vec{q})t-Z(t)} \left( \wht{S}(0,\vec{q}) + \int_0^t \!\D\tau\:
e^{\Omega(\vec{q})\tau+Z(\tau)} \wht{\eta}(\tau,\vec{q}) \right)
\EEQ
and with the abbreviations
\BEQ \label{eq:disper}
\Omega(\vec{q}) = \omega(\vec{q}) + 2\II \sum_{a=1}^d v_a \sin q_a = 2 \sum_{a=1}^d \bigl( 1 -\cos q_a + \II v_a \sin q_a \bigr) \;\; , \;\;
Z(t) = \int_0^t \!\D\tau'\: \mathfrak{z}(\tau')
\EEQ
For the computation of the two-time correlator $\wht{C}(t,t';\vec{q})$ one needs the gaussian noise correlator, along with
$\left\langle \wht{\eta}(t,\vec{q})\right\rangle=0$ and defines
\BEA
\left\langle \wht{\eta}(t,\vec{q})\wht{\eta}(t',\vec{q}')\right\rangle &=& 2T (2\pi)^d \delta(t-t') \delta(\vec{q}+\vec{q}')
\\
\left\langle \wht{S}(t,\vec{q})\wht{S}(t',\vec{q}')\right\rangle &=&  (2\pi)^d \delta(\vec{q}+\vec{q}') \wht{C}(t,t';\vec{q})
\EEA
With the further abbreviation $g(t) = e^{2Z(t)}$, one arrives at the single-time correlator
\BEQ
\wht{C}(t,\vec{q}) := \wht{C}(t,t,\vec{q})
= \frac{e^{-2\omega(\vec{q})t}}{g(t)} \left( \wht{C}(0,\vec{q}) + 2T \int_0^t \!\D\tau\: e^{2\omega(\vec{q})\tau} g(\tau) \right)
\EEQ
The spherical constraint is then recast into the condition
\BEQ \label{eq:sphercon}
\frac{1}{(2\pi)^d} \int_{\cal B} \!\D\vec{q}\: \wht{C}(t,\vec{q}) = 1
\EEQ
where ${\cal B}=[-\pi,\pi]^d$ is the Brillouin zone. In terms of the function $g(t)$,
eq.~(\ref{eq:sphercon}) takes the form of a Volterra integral equation
\BEQ \label{eq:SC}
g(t) = A(t) + 2T \int_0^t \!\D\tau\: f(t-\tau) g(\tau)
\EEQ
with the auxiliary functions \cite{Picone02}
\BEQ \label{eq:fct-aux}
f(t) := \frac{1}{(2\pi)^d} \int_{\cal B} \!\D\vec{q}\: e^{-2\omega(\vec{q}) t} = \left( e^{-4t} I_0(4t) \right)^d \;\; , \;\;
A(t) := \frac{1}{(2\pi)^d} \int_{\cal B} \!\D\vec{q}\: \wht{C}(0,\vec{q})\, e^{-2\omega(\vec{q}) t}
\EEQ
Herein, we shall admit a fully disordered initial state in the transverse directions but long-range initial correlations in the preferred direction.
{}From now on we shall take the bias as $\vec{v}=v\vec{e}_1$ for simplicity, and we shall also write $\vec{q}=(q_1,\vec{q}_{\perp})\in\mathbb{R}\otimes\mathbb{R}^{d-1}$.
The initial correlator is in the small-momentum regime
\BEQ \label{eq:iniC}
\wht{C}(0,\vec{q}) = \wht{C}(0,q_1,\vec{q}_{\perp}) = c_{\alpha} \bigl| q_1\bigr|^{\alpha}
\EEQ
with $\alpha<0$ and a positive constant\footnote{The case $\alpha=0$ of a fully disordered initial state was studied previously \cite{Godreche13}.} $c_{\alpha}>0$.
The asymptotic analysis of the constraint (\ref{eq:SC}) will be given in the appendix.
From these results it follows that the time-dependent single-time correlator $\wht{C}(t,\vec{q})$, and {\it a fortiori} the Lagrange multiplier
expressed through $g(t)$, is independent of the bias $\vec{v}$ for a fully {\em un}correlated initial state \cite{Godreche13}.
Then the behaviour of $g(t)$ is known \cite{Godr00b}.
For the initial correlations (\ref{eq:iniC}) with $\alpha<0$, this is different, as we shall see below.

On the other hand, the bias does occur in the expression of the two-time quantities, assuming here and throughout $t>s$ for notational simplicity:
\begin{itemize}
\item  For the two-time correlation function in the Fourier space one obtains:
\BEQ\label{BiasCorr}
\wht{C}(t,s;\vec{q}) = \frac{\exp\left(-\Omega(\vec{q})t - \Omega(-\vec{q})s\right)}{\sqrt{g(t) g(s)\,}\,}
\left( \wht{C}(0;\vec{q}) + 2T \int_0^s \!\D\tau\: e^{2\omega(\vec{q})\tau} g(\tau) \right)
\EEQ
It follows that in addition to the bias the two-time correlation function depends also on the initial correlator.
The effects of long-range initial correlations $\wht{C}(0;\vec{q})$ shall be of particular interest in our study.
\item The response function is obtained  by adding a small perturbation to the Hamiltonian by a magnetic-field term
$\delta{\cal H}= \sum_{\vec{r}} h(t,\vec{r})S(t,\vec{r})$. This leads to an extra term $h(t,\vec{r})$
on the right-hand side of the Langevin equation (\ref{Langeq}) and to a modified solution in Fourier space
\BEQ\label{solrep}
\wht{S}(t,\vec{q}) = e^{-\Omega(\vec{q})t-Z(t)} \left( \wht{S}(0,\vec{q}) + \int_0^t \!\D\tau\:
e^{\Omega(\vec{q})\tau+Z(\tau)}[ \wht{\eta}(\tau,\vec{q})+\wht{h}(\tau,\vec{q})] \right)
\EEQ
From the above the response function is easily found. In Fourier space one obtains
\BEQ\label{BiasResp}
\wht{R}(t,s,\vec{q})=\left.\frac{\delta\bigl\langle\wht{S}(t,\vec{q})\bigr\rangle}{\delta\wht{h}(s,\vec{q})}\right|_{\wht{h}=0}
                    =e^{-\Omega(\vec{q})(t-s)}\sqrt{\frac{g(s)}{g(t)}}
\EEQ
\end{itemize}
In these definitions, $s$ is called the {\em waiting time} and $t>s$ the {\em observation time}.

For our symmetry analysis in the following sections, we first take the continuum limit of (\ref{Langeq})
and notice that since for quenches to $T\leq T_c$, one has asymptotically
$g(t) \sim t^{\digamma}$, where $\digamma$ is given in table~\ref{tabA2}. Hence $\mathfrak{z}(t)\sim \frac{\digamma}{2}\frac{1}{t}$,
and the long-time behaviour of the model should follow from an equation of the form
\begin{align} \label{prov:digamma}
\partial_t S(t,\vec{r}) = \Delta_{\vec{r}} S(t,\vec{r}) -2\vec{v}\cdot\vec{\nabla}_{\vec{r}} S(t,\vec{r})+ \frac{\digamma}{2t} S(t,\vec{r}) + \eta(t,\vec{r})
\end{align}
This does not yet yield the from of the invariant Schr\"odinger operator $\mathscr{S}$ in table~\ref{tab1},
but the form (\ref{prov:digamma}) will appear after the change of representation explained in section~\ref{sec:generators}.
The initial long-range correlations (\ref{eq:iniC}) will become important in this asymptotic analysis.

\section{Time-space responses and correlators} \label{sec:reponse-corr}

In our analysis of meta-conformal and meta-Schr\"odinger dynamical symmetries,
we turn the coordinate axes such that the bias is along the first axis $\vec{v} = v\vec{e}_1$, which is
called the {\em preferred direction}. The other $d_{\perp}=d-1$ directions where $v_a=0$ for $a=2,\ldots,d$, are called {\em transverse directions}.

We begin with the \underline{two-time response}, which in Fourier space is given by (\ref{BiasResp}).
Working in Fourier space has the advantage that the different scaling limits to be considered
are more readily distinguished than it is possible in direct space.
The response function does not depend explicitly on the initial conditions, which only enter through the
long-time behaviour of the auxiliary function $g(t)$. For quenches to $T\leq T_c(d)$, $g(t) \simeq g_{\infty} t^{\digamma}$ will be derived in
the appendix and the values of $\digamma$ are listed in table~\ref{tabA2}. The dispersion relation becomes in momentum  space, in the limit of low momenta
\BEQ \label{Omega-bas}
\Omega(\vec{q}) = 2 \sum_{a=1}^{d} \bigl[ \big( 1 - \cos q_a\bigr) + \II v_a \sin q_a \bigr] \simeq  \bigl( q_1^2 + \vec{q}_{\perp}^2 \bigr) + 2 \II v q_1
\EEQ
with the vector decomposition $\vec{q}=\bigl( q_1, \vec{q}_{\perp}\bigr)\in\mathbb{R}^d$
into preferred and transverse components.\footnote{This formal expansion will lead to un-bounded two-point functions in direct space in the meta-conformal and
meta-Schr\"odinger cases.
A regularised form can be obtained by dualising the generators which can be shown to eliminate all spurious singularities \cite{Henkel16,Henkel20}.}
We then must consider two distinct scaling limits.
\begin{enumerate}
\item The {\em meta-conformal scaling limit} when $\tau=t-s\to\infty$ and $q_a\to 0$ such that $(t-s) q_a$ is kept fixed in all spatial directions $a=1,2,\ldots,d$.

Then $\Omega(\vec{q})\bigl(t-s\bigr)\simeq 2\II v q_1 \bigl(t -s\big) + \ldots$, up to corrections to scaling. The two-time response
in momentum space simply is
\BEQ
\wht{R}(t,s;\vec{q}) = e^{-2\II v q_1 (t-s)} \left(\frac{s}{t}\right)^{\digamma/2}
\EEQ
which in direct space then leads to the scaling form
\BEA
R(t,s;\vec{r}) &=& \frac{1}{(2\pi)^d} \int_{\mathbb{R}^d} \!\D\vec{q}\: e^{\II \vec{r}\cdot\vec{q}}\,e^{-2\II v q_1 (t-s)} \left(\frac{s}{t}\right)^{\digamma/2}
\nonumber \\
&=& \frac{1}{2\pi} \int_{\mathbb{R}} \!\D q_1\: e^{\II q_1 ( r_1 - 2v (t-s))} \frac{1}{(2\pi)^{d-1}} \int_{\mathbb{R}^{d-1}} \!\D \vec{q}_{\perp}\:
e^{\II \vec{r}_{\perp}\cdot\vec{q}_{\perp}} \left(\frac{s}{t}\right)^{\digamma/2} \nonumber \\
&=& \frac{1}{2v}\bigl( t-s\bigr)^{-d} \left(\frac{t}{s}\right)^{-\digamma/2}\:
    \delta\left( \frac{r_1}{2v (t-s)}-1\right) \delta^{(d-1)}\left( \frac{\vec{r}_{\perp}}{t-s}\right)
\EEA
where $\delta$ denotes the Dirac distribution. The cases $T<T_c(d)$ and the different regimes at $T=T_c(d)$
are merely distinguished by the value of $\digamma$, see table~\ref{tabA2}.
\item The {\em meta-Schr\"odinger scaling limit} when $\tau=t-s\to\infty$ and $q_a\to 0$ such that $(t-s) q_1$
as well as $(t-s) q_a^2$ are kept fixed for $a=2,\ldots,d$.

Then $\Omega(\vec{q})\bigl(t-s\bigr)\simeq 2\II v q_1 \bigl(t -s\big) +\vec{q}_{\perp}^2 (t-s)+ \ldots$, up to corrections to scaling.
The two-time response in momentum space is
\BEQ \label{gl:R-msch}
\wht{R}(t,s;\vec{q}) = e^{-2\II v q_1 (t-s)-\vec{q}_{\perp}^2 (t-s)}\left(\frac{s}{t}\right)^{\digamma/2}
\EEQ
which leads in direct space to the scaling form
\BEA
R(t,s;\vec{r}) &=& \frac{1}{(2\pi)^d} \int_{\mathbb{R}^d} \!\D\vec{q}\: e^{\II\vec{r}\cdot\vec{q}}\,
                   e^{-2\II v q_1 (t-s)-\vec{q}_{\perp}^2 (t-s)}\left(\frac{s}{t}\right)^{\digamma/2}
\nonumber \\
&=& \frac{1}{2\pi} \int_{\mathbb{R}} \!\D q_1\: e^{\II q_1 ( r_1 - 2v (t-s))} \frac{1}{(2\pi)^{d-1}} \int_{\mathbb{R}^{d-1}} \!\D \vec{q}_{\perp}\:
e^{\II \vec{r}_{\perp}\cdot\vec{q}_{\perp}}\, e^{-\vec{q}_{\perp}^2 (t-s)} \left(\frac{s}{t}\right)^{\digamma/2} \nonumber \\
&=& \delta\bigl( r_1 - 2v (t-s)\bigr) \bigl( 4\pi s\bigr)^{-(d-1)/2}
\left( \frac{t}{s} -1\right)^{-(d-1)/2} \left(\frac{s}{t}\right)^{\digamma/2} e^{-\vec{r}_{\perp}^2/(4(t-s))}
\nonumber \\
&=& \frac{\pi^{1/2}}{v} R^{[{\rm ms}]}(t,s)\: \delta\left( \frac{1}{2 v}\frac{r_1}{t-s} - 1\right) \exp\left[ -\frac{\vec{r}_{\perp}^2}{4(t-s)} \right]
\EEA
with the auto-response $R^{[{\rm ms}]}(t,s) = \bigl( 4\pi s\bigr)^{-(d+1)/2} \left( \frac{t}{s} -1\right)^{-(d+1)/2} \left(\frac{t}{s}\right)^{-\digamma/2}$
which is of the same form as for un-biased case \cite{Picone02}, but with different scaling dimensions.
Again, the several situations at $T\leq T_c(d)$ are merely distinguished by the value of $\digamma$, see table~\ref{tabA2}.
\end{enumerate}

\begin{figure}[tb]
\begin{center}
\includegraphics[width=.3\hsize]{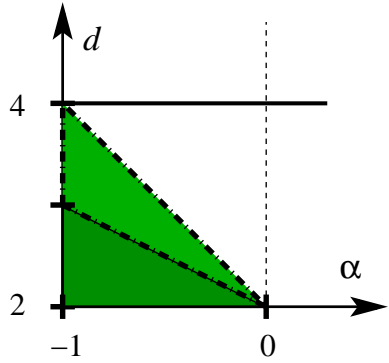}
\end{center}
\caption[fig2]{\small Regions in $(\alpha,d)$-parameter space, with $-1<\alpha<0$ and $d>2$, for a quench onto $T=T_c(d)$.
In the shaded regions, which correspond to the regimes $2<d<2-\alpha$ and $2-\alpha<d<4$ respectively, 
the initial long-range correlations (\ref{eq:iniC}) are more relevant than the thermal noise.
 \label{fig2}}
\end{figure}

Next, we analyse the \underline{two-time correlator}, which in Fourier space is given by (\ref{BiasCorr}).
The relative relevance of the two terms coming from initial noise and thermal noise,
respectively, must be analysed. With the low-energy limit (\ref{Omega-bas}) of the dispersion relation, we have using (\ref{gl:A.14}) and the initial condition (\ref{eq:iniC})
\BEA
\lefteqn{\hspace{-0.3cm}\wht{C}(t,s;\vec{q}) = \frac{\exp\bigl[ - 2\II v q_1 (t-s) - \vec{q}_{\perp}^2(t+s)\bigr]}{g_{\infty} \bigl( t s\bigr)^{\digamma/2}}
\left( c_{\alpha} \bigl| q_1\bigr|^{\alpha} + 2T \int_0^s \!\D \tau\: e^{2\vec{q}_{\perp}^2 \tau}
\left( \frac{\mathscr{A}_0}{M^2}\delta(\tau) + g_{\infty} \tau^{\digamma} \right) \right)}
\nonumber \\
&\simeq& \frac{\exp\bigl[ - 2\II v q_1 (t-s) - \vec{q}_{\perp}^2(t+s)\bigr]}{\bigl( t s\bigr)^{\digamma/2}}
         \frac{c_{\alpha}}{g_{\infty}} \bigl| q_1\bigr|^{\alpha}
\nonumber \\
& & + \frac{\exp\bigl[ - 2\II v q_1 (t-s) - \vec{q}_{\perp}^2(t+s)\bigr]}{\bigl( t s\bigr)^{\digamma/2}} 2T \int_0^s \!\D \tau\:
      e^{2\vec{q}_{\perp}^2 \tau} \tau^{\digamma}
\label{Ctilde} \\
&=& \frac{\exp\bigl[ - 2\II v q_1 (t-s) - \vec{q}_{\perp}^2(t+s)\bigr]}{\bigl( t s\bigr)^{\digamma/2}}
\left( \frac{c_{\alpha}}{g_{\infty}} \bigl| q_1\bigr|^{\alpha} +
\frac{2T}{1+\digamma} s^{1+\digamma} {}{_{1}F_{1}}\bigl( 1+\digamma, 2+\digamma;  2\vec{q}_{\perp}^2 s\bigr) \right)
\nonumber
\EEA
where $_1F_1$ is a Kummer function and \cite[(13.2.1)]{Abra65} was used.
In the first line, we recognise that for $\alpha<0$, the singular contribution $\sim \mathscr{A}_0\delta(\tau)$ merely contributes a
correction to the leading scaling. As to the two scaling limits, in the meta-conformal limit, we have $\vec{q}_{\perp}^2s \to 0$
and in the meta-Schr\"odinger limit, $\vec{q}_{\perp}^2s$
is finite. Therefore, in (\ref{Ctilde}), since the dynamical exponent $z_{\|}=1$, the first term (related to the initial correlator) is more relevant, if
\BEQ \label{rel-cond}
\alpha <  -1-\digamma
\EEQ
Using table~\ref{tabA2}, we can distinguish the following situations for $T\leq T_c(d)$:
\begin{enumerate}
\item For $T<T_c(d)$, we should have $\alpha< d-2$. Since $d>2$ is needed to have a non-vanishing $T_c(d)$,
but $\alpha$ is always negative, that condition is automatically satisfied. 
Hence the thermal noise term is irrelevant in the entire ordered phase.
\item For $T=T_c(d)$ and $2<d<2-\alpha$, the condition (\ref{rel-cond}) leads to $\alpha<\alpha/2$, hence $\alpha<0$.
Since this is always the case, in a critical system with $2<d<2-\alpha$ the thermal noise term is always irrelevant.
\item For $T=T_c(d)$ and $2-\alpha < d<4$, we find from (\ref{rel-cond}) that $\alpha< 1-d/2$.
Only for these values of $\alpha$, which means that the initial
spatial correlator should decay more slowly that $C_n(0)\sim |n|^{-(1+\alpha)}$,
the initial noise is relevant and the thermal noise generates a correction to scaling. On the other hand, if $0>\alpha>1-d/2$,
then the thermal noise is relevant and the initial noise generates a correction to scaling.\footnote{This case is the analogue of what is found
in the biased $1D$ Glauber-Ising model at $T=0$ with long-range initial correlations \cite{Henkel19a}.}
\item For $T=T_c(d)$ and $d>4$, we have $\digamma =0$, hence $\alpha<-1$.
This is impossible which means that the thermal noise is always relevant and
that now  the initial noise generates a correction to the leading scaling.
\end{enumerate}
Figure~\ref{fig2} shows in the $(\alpha,d)$-parameter space as shaded those regions where the initial noise is more relevant than the thermal noise, for
the directed spherical model quenched onto its critical point $T=T_c(d)$.
For short-ranged initial correlations with $\alpha=0$, the thermal noise is always more relevant \cite{Godreche13}.
The relevance of certain long-ranged initial correlations is certainly induced by fluctuations, since this does not occur in the mean-field regime at $d>4$.

These results have the following consequences for the scaling form of the two-time correlator:
\begin{enumerate}
\item If the initial noise is relevant, the two-point correlator is in momentum space
\BEQ \label{gl:C-ini}
\wht{C}(t,s;\vec{q}) = \frac{c_{\alpha}}{g_{\infty}} \frac{\exp[-2\II v q_1 (t-s) - \vec{q}_{\perp}^2 (t+s)]}{(t s)^{\digamma/2}}
\bigl| q_1\bigr|^{\alpha}
\EEQ
but the form in direct space does depend on the scaling limit to be taken.\\
{\bf A.} In the \underline{meta-conformal limit}, the term $\vec{q}_{\perp}^2 (t+s)\to 0$,
so that the scaling form in time-space of the two-point correlator is
\BEA
C(t,s;\vec{r}) &=& \frac{1}{(2\pi)^d} \int_{\mathbb{R}^d} \!\D\vec{q}\: e^{\II \vec{q}\cdot\vec{r}}
\frac{c_{\alpha}}{g_{\infty}} \frac{\exp[-2\II v q_1 (t-s)]}{(t s)^{\digamma/2}}
\bigl| q_1\bigr|^{\alpha} \nonumber \\
&=& \frac{1}{2\pi} \int_{\mathbb{R}} \!\D q_1\: \frac{\exp[ \II q_1\bigl( r_1 - 2 v (t-s)\bigr)]}{(t s)^{\digamma/2}}
\bigl| q_1\bigr|^{\alpha} \frac{c_{\alpha}}{g_{\infty}}
\delta^{(d-1)}(\vec{r}_{\perp}) \nonumber \\
&=& C_{(0)}\: \bigl( t s\bigr)^{-\digamma/2}\: \bigl| r_1 - 2 v (t-s)\bigr|^{-\alpha-1}  \delta^{(d-1)}(\vec{r}_{\perp})
\EEA
with the normalisation constant $C_{(0)}:= \frac{c_{\alpha}}{2\pi g_{\infty}} \int_{\mathbb{R}}\!\D q_1\: e^{\II q_1} \bigl| q_1\bigr|^{\alpha}$
(to be interpreted as a distribution \cite{Gelf64}). \\
{\bf B.} In the \underline{meta-Schr\"odinger limit}, we rather have
\BEA
C(t,s;\vec{r}) &=& \frac{1}{(2\pi)^d} \int_{\mathbb{R}^d} \!\D\vec{q}\: e^{\II \vec{q}\cdot\vec{r}} \frac{c_{\alpha}}{g_{\infty}}
\frac{\exp[-2\II v q_1 (t-s)- \vec{q}_{\perp}^2 (t+s)]}{(t s)^{\digamma/2}}
\bigl| q_1\bigr|^{\alpha} \nonumber \\
&=& \frac{1}{2\pi} \int_{\mathbb{R}} \!\D q_1\:
    \frac{\exp[ \II q_1\bigl( r_1 - 2 v (t-s)\bigr)]}{(t s)^{\digamma/2}} \bigl| q_1\bigr|^{\alpha} \frac{c_{\alpha}}{g_{\infty}}
    \frac{1}{(2\pi)^{d-1}} \int_{\mathbb{R}^{d-1}} \!\D\vec{q}_{\perp}\: e^{\II \vec{q}_{\perp}\cdot\vec{r}_{\perp}- \vec{q}_{\perp}^2 (t+s)}
\nonumber \\
&=& C_{[0]}\:  \bigl( t s\bigr)^{-\digamma/2}\:
\bigl| r_1 - 2 v (t-s)\bigr|^{-\alpha-1}\: \bigl(t+s\bigr)^{-(d-1)/2}\: \exp\left[-\frac{1}{4} \frac{\vec{r}_{\perp}^2}{t+s}\right]~~~
\EEA
and several constants were absorbed into $C_{[0]}$.

These two cases arise in the spherical model in the entire ordered phase $T<T_c(d)$, and at criticality $T=T_c(d)$
if either $d<2-\alpha$ or else for $2-\alpha<d<4$ if $\alpha<1-d/2$.
\item If the thermal noise is relevant, however,
\BEA
\wht{C}(t,s;\vec{q}) &=& \frac{\exp\bigl[ - 2\II v q_1 (t-s)-\vec{q}_{\perp}^2(t+s)\bigr]}{\bigl( t s\bigr)^{\digamma/2}}
                         2T_c(d) \int_0^s \!\D \tau\: e^{2\vec{q}_{\perp}^2 \tau} \tau^{\digamma}
\nonumber \\
&=& \frac{2T_c(d)}{1+\digamma}  s\left(\frac{s}{t}\right)^{\digamma/2} e^{- 2\II v q_1 (t-s) - \vec{q}_{\perp}^2(t+s)}
 {}{_{1}F_{1}}\bigl( 1+\digamma, 2+\digamma;  2\vec{q}_{\perp}^2 s\bigr)~~~  \label{gl:C-therm}
\EEA
reduces to the un-biased correlator, up to a small damping term.
This is the situation which has been discussed in detail in the litt\'erature \cite{Godreche13}.
This case occurs in the critical spherical model ($T=T_c(d)$) for either $d>4$ or else for $2-\alpha<d<4$ if $\alpha>1-d/2$
and notably for all dimensions $d>2$ for short-ranged initial correlations where $\alpha=0$.
\end{enumerate}

Our main result on the directed spherical model, subject to the long-ranged initial correlations (\ref{eq:iniC}) in the preferred direction, is as follows:
there can exist, beyond the expected diffusive time-space scaling with dynamical exponent $z=2$, for larger spatial distances
a further ballistic dynamical scaling regime. Then the dynamical exponent is either isotropically $z=1$ in the case of a meta-conformal scaling limit or
$z_{\|}=1$ only in the preferred direction in the case of a meta-Schr\"odinger scaling limit.
These regimes arise (i) in the coexistence phase $T<T_c(d)$ and (ii) at criticality $T=T_c(d)$
in the $(\alpha,d)$ regions indicated in figure~\ref{fig2}.

\section{Representations in Fourier space} \label{sec:generators}

In order to obtain the representations of meta-conformal \cite{Henkel19a} and meta-Schr\"odinger algebras \cite{Stoimenov22} in Fourier space
one must transform the quasi-primary scaling-operator as well as the generators of meta-conformal and meta-Schr\"odinger algebras.
For simplicity we use one `parallel' or preferred direction $x\to q_{\|}=:q $ and one `perpendicular' or transverse direction $y\to q_{\perp}=:p$
in Fourier space. For example, for the the meta-Schr\"odinger algebra the fields are transformed as follows
\BEQ \label{Fourierimm}
\Phi(t,x,y)=\frac{1}{4\pi^2}\int_{\R^2}\!\D q \D p\: e^{\II(xq+yp)}\wht{\Phi}(t,q,p) \;\; , \;\;
\wht{\Phi}(t,q,p)= 
\int_{\R^2} \!\D x \D y\:  e^{-\II(xq+yp)}{\Phi}(t,q,p)
\EEQ
which provides the following correspondence between simplest differential operations\footnote{More complicated operators
follow from the more simple ones. For example
$\vec{r}^2\partial_{\vec{r}}=\vec{r}\cdot(\vec{r}\partial_{\vec{r}})\to \II\partial_{\vec{q}}\cdot(-\vec{q}\partial_{\vec{q}}-d)$,
where $d$ is the number of the space dimensions.}
\BEQ
\partial_x\to \II q, \quad x\to \II\partial_q, \quad  x^2\to -\partial_q^2, \quad \partial_x^2\to -q^2, \quad x\partial_x\to -q\partial_q-1\nonumber
\EEQ
Now we are able to write the generators of meta-conformal and meta-Schr\"odinger algebra\footnote{Here we shall work with case $\alpha=0$
for both algebras, see \cite{Henkel19a,Stoimenov22}.}, in Fourier space, that is acting on $\wht{\Phi}$.

\subsection{The meta-conformal algebra in Fourier space}

In order to find the $n$-point functions of quasi-primary scaling operators, the generators of the finite-dimensional sub-algebras are required.

\subsubsection{Standard representation}
\BEA
X_{-1} & = & -\partial_t, \qquad X_0 = -t\partial_t+q\partial_q-\delta+1\nonumber\\
X_1    & = & - t^2\partial_t+\II\beta q\partial^2_q+2tq\partial_q-2\II(\gamma-\beta)\partial_q-2t(\delta-1)
\nonumber\\
Y_{-1} & = & -\II q, \qquad Y_0 = -\II tq+\beta q\partial_q-(\gamma-\beta)\nonumber\\
Y_1    & = & -\II t^2q+2\beta tq\partial_q-2\II\beta(\gamma-\beta)\partial_q+\II\beta^2q\partial^2_q-2t(\gamma-\beta).\label{metaconffourier}
\EEA
Here, and in what follows, all generators are throughout specified for one preferred and one transverse dimension. Meta-conformal generators do not contain
the transverse momentum $p$.  The scaling dimension $\delta$ and the rapidity $\gamma$ characterise the scaling-operators on which these generators act.

\subsubsection{Ageing representation}
In contexts of physical ageing, it is often necessary to change the representation into
$X_n \mapsto e^{\xi\ln t} X_n e^{-\xi\ln t}$ and $Y_n \mapsto e^{\xi\ln t} Y_n e^{-\xi\ln t}$, where the generators now read \cite{Stoimenov22}
\BEA
X_{-1} & = & -\partial_t+\xi/t, \qquad X_0 = -t\partial_t+q\partial_q-\delta+\xi +1\nonumber\\
X_1    & = & - t^2\partial_t+\II\beta q\partial^2_q+2tq\partial_q-2\II(\gamma-\beta)\partial_q-(2\delta-\xi-2)t
\nonumber\\
Y_{-1} & = & -\II q, \qquad Y_0 = -\II tq+\beta q\partial_q-(\gamma-\beta)\nonumber\\
Y_1    & = & -\II t^2q+2\beta tq\partial_q-2\II\beta(\gamma-\beta)\partial_q+\II\beta^2q\partial^2_q-2t(\gamma-\beta)\label{agemetaconffourier}
\EEA
Here, the scaling-operators are also characterised by $\xi$, in addition  
to\footnote{An older convention \cite{Henkel06a} insisted on keeping the dilatation generator $X_n$ unchanged and reduces to the present one under the formal
substitution $\delta\mapsto \delta-\xi$. Then specific values of the exponents $\delta$, $\xi$ in models are different from what is quoted in the litt\'erature 
\cite{Henkel10}.} $\delta,\gamma$.

Physical considerations beyond pure algebra will decide which representation must be used, see section~\ref{sec:comparaison}. The invariant Schr\"odinger
operator $\mathscr{S}$ of the standard representation becomes $\lap{\mathscr{S}}=e^{\xi \ln t}\mathscr{S}e^{-\xi \ln t}=\mathscr{S}-\xi/t$ which is exactly the
kind of additional term arising in the spherical model equation of motion (\ref{prov:digamma}).

\subsection{The meta-Schr\"odinger algebra in Fourier space}

Once more, we consider quasi-primary scaling operators.

\subsubsection{Standard representation}
\BEA
X_{-1} & = & -\partial_t, \qquad X_0 = -t\partial_t+q\partial_q+\frac{p}{2}\partial_p-\delta+\frac{3}{2}\nonumber\\
X_1    & = & - t^2\partial_t+\II\beta q\partial^2_q+2tq\partial_q-2\II(\gamma-\beta)\partial_q+tp\partial_p+\frac{{\cal M}}{2}\partial^2_p-2t(\delta-3/2)
\nonumber\\
Y_{-1}^{(\|)} & = & -\II q, \qquad Y_0^{(\|)} = -\II tq+\beta q\partial_q-(\gamma-\beta)\nonumber\\
Y_1^{(\|)}    & = & -\II t^2q+2\beta tq\partial_q-2\II\beta(\gamma-\beta)\partial_q+\II\beta^2q\partial^2_q-2t(\gamma-\beta)\nonumber\\
Y_{-\demi}^{(\perp)} & = & -\II p, \qquad Y_{\demi}^{(\perp)}=-\II t p-\II {\cal M}\partial_{p}, \qquad M_0= -{\cal M}\label{metaschfourier}
\EEA
Again, we concentrate on the case of one preferred and $d_{\perp}=1$ transverse dimensions, also labelled as $\|$ and $\perp$, respectively.
Spatial rotation-invariance allows a trivial generalisation to $d_{\perp}\geq 1$ transverse dimensions.
Notice that the transverse momentum $p$ not only appears in the generators $X_n$ but also leads to the appearance of the additional generators
$Y_{\pm\demi}^{(\perp)}$ and $M_0$.

\subsubsection{Ageing representation}
The equivalent ageing representation, with $X_n \mapsto e^{\xi\ln t} X_n e^{-\xi\ln t}$ etc., has the generators \cite{Stoimenov22}
\BEA
X_{-1} & = & -\partial_t+\xi/t, \qquad X_0 = -t\partial_t+q\partial_q+\frac{p}{2}\partial_p-\delta+\xi+\frac{3}{2}\nonumber\\
X_1    & = & - t^2\partial_t+\II\beta q\partial^2_q+2tq\partial_q-2\II(\gamma-\beta)\partial_q+tp\partial_p+\frac{{\cal M}}{2}\partial^2_p-(2\delta-\xi-3)t
\nonumber\\
Y_{-1}^{(\|)} & = & -\II q, \qquad Y_0^{(\|)} = -\II tq+\beta q\partial_q-(\gamma-\beta)\nonumber\\
Y_1^{(\|)}    & = & -\II t^2q+2\beta tq\partial_q-2\II\beta(\gamma-\beta)\partial_q+\II\beta^2q\partial^2_q-2t(\gamma-\beta)\nonumber\\
Y_{-\demi}^{(\perp)} & = & -\II p, \qquad Y_{\demi}^{(\perp)} =-\II t p-\II {\cal M}\partial_{p}, \qquad M_0= -{\cal M}\label{agemetaschfourier}
\EEA
Again, the resulting Schr\"odinger operator $\lap{\mathscr{S}}=e^{\xi \ln t}\mathscr{S}e^{-\xi \ln t}=\mathscr{S}-\xi/t$
should be compared with the spherical model equation of motion (\ref{prov:digamma}).

\section{Covariant two-point functions in Fourier space} \label{sec:cov-2point}

In Fourier space one has the following expression for the covariant two-point functions
\BEQ
F = F^{[2]}(t_a,t_b,q_a,q_b,p_a,p_b)=\bigl\langle \wht{\Phi}_a(t_a,q_a,p_a)\wht{\Phi}_b(t_b,q_b,p_b)\bigr\rangle
\EEQ
where the scaling operators $\wht{\Phi}_a$ are quasi-primary,
i.e. transform covariantly with respect to the maximal finite-dimensional sub-algebra of the meta-conformal algebra
(\ref{metaconffourier}, \ref{agemetaconffourier}) or of the meta-Schr\"odinger algebra (\ref{metaschfourier}, \ref{agemetaschfourier}).
The Ward identities immediately follow using the transformation terms already included in the generators such that the $n$-point functions follow in a
standard way by solving the associated system if differential equations.

When it comes to write down the Ward identities in Fourier space, for correlators with spatial translation-invariance, the identity
\BEA
& & \int_{\mathbb{R}^n} \!\D x_1 \ldots \D x_n\: e^{-\II x_1 q_1 - \ldots - \II x_n q_n}  F^{[n]}(x_1 - x_n, \ldots, x_{n-1}-x_n) \\
&=& \int_{\mathbb{R}^n} \!\D x_1 \ldots \D x_n\: e^{-\II (x_1-x_n) q_1 - \ldots -\II (x_{n-1}-x_n) q_{n-1} -\II x_n(q_1 + \ldots + q_{n-1} + q_n)}
F^{[n]}(x_1 - x_n, \ldots, x_{n-1}-x_n) \nonumber \\
&=& 2\pi \delta(q_1+\ldots +q_n) \int_{\mathbb{R}^n} \!\D \eta_1 \ldots \D \eta_{n-1}\: e^{-\II \eta_1 q_1 - \ldots -\II \eta_{n-1} q_{n-1} }
F^{[n]}(\eta_1, \ldots, \eta_{n-1}) \nonumber
\EEA
is sometimes useful.

\subsection{Meta-conformal covariant two-point functions}

We begin with the standard representation.
Covariance under the time- and space-translations $X_{-1}, Y_{-1}$  gives, with $\tau=t_a-t_b$ and the Dirac function $\delta(q)$
\BEQ \label{Wardtranslations}
F = F^{[2]}(t_a,t_b,q_a,q_b)= \delta(q_a+q_b)\,f(\tau, q_a)
\EEQ
Next, co-variance under the generator of dynamical scaling $X_0$ implies
\BEQ \label{dynmetaconf}
\bigl(\tau\partial_{\tau}-q_a\partial_{q_a}+\delta_a+\delta_b-1\bigr)f(\tau,q_a)=0
\EEQ
Co-variance under the generator $Y_0$ gives:
\BEQ
\bigl(\II \tau q_a-\beta_a q_a\partial_{q_a}+\gamma_a+\gamma_b-\beta_b\bigr)f(\tau,q_a)=0 ~~~~\mbox{\rm\small and}~~~~
\beta_a=\beta_b=\beta. \label{covy0}
\EEQ
Carrying out the integration with respect to $q_a$ and inserting the result into (\ref{dynmetaconf}) leads to
\BEQ
f(\tau,q_a) = F_0\,\tau^{-\delta_a-\delta_b+(\gamma_a+\gamma_b)/\beta} q_a^{{(\gamma_a+\gamma_b)}/{\beta}-1} \exp\left({\frac{\II}{\beta}\tau q_a}\right)
\label{mataconftwopointmedium}
\EEQ
Finally, the covariance under $X_1$ and $Y_1$ merely gives the constraints
\BEQ \label{condx1}
\delta_a=\delta_b=\delta \;\;,\;\;  \gamma_a=\gamma_b=\gamma
\EEQ
and the final result for the two-point function covariant under representation (\ref{metaconffourier})
of meta-conformal algebra is, together with the constraints (\ref{condx1})
\BEQ \label{metaconftwopointfinal}
F^{[2]}(t_a,t_b,q_a,q_b)=F_0\,\delta(q_a+q_b)\,\bigl(t_a-t_b\bigr)^{-2\delta+2\gamma/\beta}q_a^{{2\gamma}/{\beta}-1}
\exp\left({\frac{\II q_a}{\beta}(t_a-t_b)}\right)
\EEQ
where $F_0$ is an undetermined normalisation constant.

Now, we look for the form of two-point function covariant under ageing representation (\ref{agemetaconffourier})
of meta-conformal algebra. Here time-translations-invariance is explicitly broken via the time-dependence of the generator
$X_{-1}=-\partial_t+\xi/t$. Since this modified generator is equivalent to the standard time-translations
$X_{-1} = e^{\xi \ln t} \bigl( -\partial_t\bigr) e^{-\xi\ln t}$,
the co-variance under $X_{-1}$ leads to the following equation for the  two-point function $F=F(\tau,u, t_b, q_a,q_b)$, after a change of
variables $\tau=t_a-t_b$ and $u=t_a/t_b$
\BEQ \label{ucondition}
\bigl((u-1)u\partial_u+\xi_a+u\xi_b\bigr)F(\tau,u,q_a,q_b)=0
\EEQ
Integrating (\ref{ucondition}), the $u$-dependence is determined
\BEQ  \label{udependence2}
F(\tau, u, q_a,q_b)= u^{\xi_a}(u-1)^{-\xi_a-\xi_b} \delta(q_a+q_b)\, f(\tau,q_a)
\EEQ
where spatial translation-invariance is also taken into account. Herein, the function $f(\tau,q_a)$ is the same as studied above, with the only difference that
the condition of dilation-covariance now reads
\BEQ \label{dynmetaconf-xi}
\bigl(\tau\partial_{\tau}-q_a\partial_{q_a}+\delta_a+\delta_b-\xi_a-\xi_b-1\bigr)f(\tau,q_a)=0
\EEQ
The other equations are unchanged and covariance under $X_1$ and $Y_1$ reproduces (\ref{condx1}).
Consequently, the co-variant two-point function under the ageing representation (\ref{agemetaconffourier})
of the meta-conformal algebra is given by
\BEQ \label{agemetaconftwopointfinal}
F^{[2]}(t_a,t_b,q_a,q_b)=F_0\,\delta(q_a+q_b)\,t_a^{\xi_a} t_b^{\xi_b}
                   \bigl(t_a-t_b\bigr)^{-\delta_a-\delta_b+2\gamma/\beta}q_a^{{2\gamma}/{\beta}-1}\exp\left(\frac{\II q_a}{\beta}(t_a-t_b)\right)
\EEQ
with the constraints (\ref{condx1}).

\subsection{Meta-Schr\"odinger covariant functions}

\subsubsection{Covariant two-point functions}
We proceed in the same manner writing the Ward identities for the two-point function
\BEQ
F^{[2]}(t_a,t_b,q_a,q_b,p_a,p_b)=\bigl\langle \wht{\Phi}_a(t_a,q_a,p_a)\wht{\Phi}_b(t_b,q_b,p_b)\bigr\rangle\nonumber
\EEQ
Co-variance under $X_{-1}, M_0, Y_{-1}^{(\|)},Y_{-\demi}^{(\perp)}$ taken from (\ref{metaschfourier}) gives (again with $\tau=t_a-t_b$)
\BEQ \label{SWardtranslations}
F^{[2]}(t_a,t_b,q_a,q_b,p_a,p_b)=\delta({\cal M}_a+{\cal M}_b)\delta_1(q_a+q_b)\delta(p_a+p_b)\,f(\tau, q_a,p_a)
\EEQ
Next, co-variance under the generator of dynamical scaling $X_0$ produces
\BEQ \label{dynmetasch}
\left(\tau\partial_{\tau}-q_a\partial_{q_a}-\frac{p_a}{2}\partial_{p_a}+\delta_a+\delta_b+\frac{3}{2}\right)f(\tau,q_a,p_a)=0
\EEQ
Covariance under the generator $Y_0^{(\|)}$ gives:
\BEQ
\bigl(\II \tau q_a-\beta_a q_a\partial_{q_a}+\gamma_a+\gamma_b-\beta_b\bigr)f(\tau,q_a,p_a)=0 ~~~~\mbox{\rm\small and}~~~~
\beta_a=\beta_b=\beta \label{covschy0}
\EEQ
Furthermore, covariance under $Y_{\demi}^{(\perp)}$ implies
\BEQ\label{galcovariance}
\bigl(\tau p_a+{\cal M}_a\partial_{p_a}\bigr)f(\tau,q_a,p_a)=0.
\EEQ
This last condition can be integrated with respect to $p_a$. The result is
\BEQ\label{pdependence}
f(\tau,q_a,p_a)=f_1(\tau,q_a)\,e^{-p_a^2 \tau/(2{\cal M}_a)}.
\EEQ
Substituting this into (\ref{dynmetasch},\ref{covschy0}), we obtain
\BEQ
f(\tau,q_a,p_a)= F_0\,\tau^{-\delta_a-\delta_b+(\gamma_a+\gamma_b)/\beta+\demi}q_a^{{(\gamma_a+\gamma_b)}/{\beta}-1}
\exp\left(\frac{\II}{\beta}\tau q_a-\frac{p_a^2}{2{\cal M}_a}\tau\right) 
\EEQ
Finally the covariance under $X_1$ and $Y_1^{(\|)}$ reproduces (\ref{condx1}).
This leads to the following form of covariant two-point function under the standard representation (\ref{metaschfourier})
\BEA
F^{[2]}(t_a,t_b,q_a,q_b,p_a,p_b) & = & F_0\, \delta({\cal M}_a+{\cal M}_b)\delta(q_a+q_b)\delta(p_a+p_b) \label{twopointmetaschfinal} \times\\
      & & \times  \bigl(t_a-t_b\bigr)^{-2\delta+2\gamma/\beta+\demi}q_a^{{2\gamma}/{\beta}-1}
                  \exp\left( \frac{\II q_a}{\beta}(t_a-t_b)-\frac{p_a^2}{2{\cal M}_a}(t_a-t_b)\right) \nonumber
\EEA
with the constraint (\ref{condx1}).

Covariance under the ageing representation (\ref{agemetaschfourier}) again leads to an additional dependence on $u=t_a/t_b$, through the modified
generator $X_{-1}=-\partial_t+\xi/t$. The corresponding Ward identity is analogous to (\ref{ucondition}) for the  meta-conformal case.
The other Ward identities remain unchanged and also reproduce once more (\ref{condx1}), with the exception of dilation-invariance.
We simply quote the final result
\BEA
\lefteqn{F^{[2]}(t_a,t_b,q_a,q_b,p_a,p_b)  = F_0\,\delta({\cal M}_a+{\cal M}_b)\delta(q_a+q_b)\delta(p_a+p_b)\times} \label{twopointmetaschfinal2}\\
& & \times  t_a^{\xi_a} t_b^{\xi_b} \bigl(t_a-t_b\bigr)^{-\delta_a-\delta_b+2\gamma/\beta+d_{\perp}/2}
    q_a^{{2\gamma}/{\beta}-1}
    \exp\left( \frac{\II q_a}{\beta}(t_a-t_b)-\frac{p_a^2}{2{\cal M}_a}(t_a-t_b)\right) \nonumber
\EEA
with the constraint (\ref{condx1}). Assuming rotation-invariance in the transverse directions,
we have also restored an arbitrary transverse dimension $d_{\perp}=d-1\geq 1$, for later reference.


\section{Comparison with the spherical model} \label{sec:comparaison}

The comparison of generic forms extracted from the representations of the meta-conformal and meta-Schr\"odinger algebras with the explicit results of
the spherical model is simplified in momentum space, where the various scaling limits needed can be formulated more readily. It is also important to recall
that the basic $n$-point functions of meta-conformal invariance are correlators $\langle \Phi\Phi\cdots\rangle$
\cite{Henkel16,Henkel20}, whereas the basic $n$-point functions of meta-Schr\"odinger invariance are responses
$\left.\frac{\delta \langle \Phi\cdots\rangle}{\delta h\cdots}\right|_{h=0}=\langle \Phi \cdot \wit{\Phi}\cdots\rangle$ \cite{Stoimenov22},
and where $\wit{\Phi}$ is the response operator associated to the conjugate field $h$ in a perturbation $\delta{\cal H}=h \Phi$, in the context of
Janssen-de Dominicis theory, see e.g. \cite{Taeu14}.

\subsection{Responses and meta-Schr\"odinger invariance}

Meta-Schr\"odinger invariance predicts the form of the two-point {\em response},
written as a correlator of the order parameter $\Phi$ with a response operator $\wit{\Phi}$, and both assumed quasi-primary, such that
$R=\bigl\langle \Phi \wit{\Phi}\bigr\rangle$.
The comparison must be done between the explicit spherical model result (\ref{gl:R-msch}) and the prediction (\ref{twopointmetaschfinal2}) of meta-Schr\"odinger
invariance. Comparing the exponential factors, we identify
\BEQ
\frac{1}{\beta} = - 2 v \;\; , \;\; {\cal M}_1 = \demi
\EEQ
and comparison of the algebraic prefactors further leads to, using also (\ref{condx1})
\BEQ
\xi_1 = - \xi_2 = \frac{\digamma}{2} \;\; , \;\; \delta_1 = \delta_2 = \demi +\frac{d_{\perp}}{4} = \frac{d+1}{4}
\;\; , \;\; \frac{\gamma_1}{\beta} = \frac{\gamma_2}{\beta} = \demi
\EEQ
and  the values of $\digamma$ are listed in table~\ref{tabA2}. We finally have the scaling dimensions $\delta=\delta_1$ of the order parameter,
$\wit{\delta}=\delta_2$ of the response operator, the second scaling dimensions $\xi=-\wit{\xi}=\xi_1$ and the rapidities $\gamma/\beta=\wit{\gamma}/\beta=\demi$.
These identifications hold for all temperatures $T\leq T_c(d)$ and all dimensions $d>2$ and it is satisfying that always 
$\delta>0$ and $\wit{\delta}>0$. It follows that {\em the scaling two-response function in the directed spherical model perfectly 
reproduces the prediction of meta-Schr\"odinger invariance}. In order to achieve this, it is enough to work at sufficiently large distances in the preferred
direction such that the scaling limit (\ref{gl:metaSchr}) holds.

\subsection{Correlators and meta-conformal invariance}

Meta-conformal invariance predicts the form of the two-point {\em correlator} $C=\bigl\langle \Phi\Phi\bigr\rangle$
of a quasi-primary scaling operator $\Phi$. The comparison must be done between the explicit spherical model result (\ref{Ctilde})
and the prediction (\ref{agemetaconftwopointfinal}) of meta-conformal invariance.
Since the spherical model result (\ref{Ctilde})
has two contributions, one from the initial noise and one from the thermal noise,
two cases must be distinguished, depending on the relevance of both terms, according to the criterion (\ref{rel-cond}).

In the first case, when the \underline{initial noise is relevant}, it is enough to compare (\ref{gl:C-ini})
with the meta-conformal prediction (\ref{agemetaconftwopointfinal}).
{}From the exponential factors and the momentum-dependence, we read off
\BEQ
\frac{1}{\beta} = - 2 v \;\; , \;\; \frac{2\gamma_1}{\beta} = 1 + \alpha
\EEQ
and comparison of the exponents of the time-dependent factors leads to
\BEQ
\xi_1 = \xi_2 = -\frac{\digamma}{2} \;\; , \;\; \delta_1 + \delta_2 - \frac{2\gamma}{\beta} = 0
\EEQ
which gives for the scaling dimension and the rapidity
\BEQ
\delta_1 = \delta_2 = \frac{1+\alpha}{2} >0 \;\; ,\;\; \frac{\gamma_1}{\beta} = \frac{\gamma_2}{\beta} = \frac{1+\alpha}{2}
\EEQ
It is satisfying that the scaling dimensions $\delta_1=\delta_2$ come out to be positive.
We also find that the second scaling dimension $\xi_1=\xi_2=-\digamma/2>0$ does not vanish.
Its value depends in general on the initial correlator through $\alpha$,
with the only exception at $T=T_c(d)$ and $2-\alpha<d$.
This means that {\em in the directed spherical model with relevant long-ranged initial conditions, a meta-conformal
scaling regime exists} either for quenches to $T<T_c(d)$ for all $\alpha<0$ or else for quenches to
$T=T_c(d)$ where $d<2-\alpha$ and $-1<\alpha<0$
or for $2-\alpha<d<4$ and $\alpha<1-d/2$.

In the second case, the \underline{thermal noise is relevant} and we now compare (\ref{gl:C-therm}) with the prediction
(\ref{agemetaconftwopointfinal}), where we take into account that $\vec{q}_{\perp}^2s\to 0$
in the meta-conformal scaling limit. As before, comparison of the exponential factors and the momentum-dependence would lead to
$1/\beta=-2v$ and $2\gamma/\beta=1$. Then, identifying the power-law terms in $t$ and $s$, we would read off
\BEQ
\xi_1 = - \frac{\digamma}{2} \;\; , \;\; 2\delta =1
\EEQ
which would imply $\delta_1 = \delta_2=\demi$ and $\xi_2 = 1-\xi_1 = 1+\digamma/2$.
The finding $\xi_1\ne \xi_2$ is incompatible with a correlator of two identical scaling operators.
This means that {\em in this second case, a meta-conformal scaling regime does not exist}.

\section{Conclusions} \label{sec:concludere}

\begin{figure}[tb]
\begin{center}
\includegraphics[width=.5\hsize]{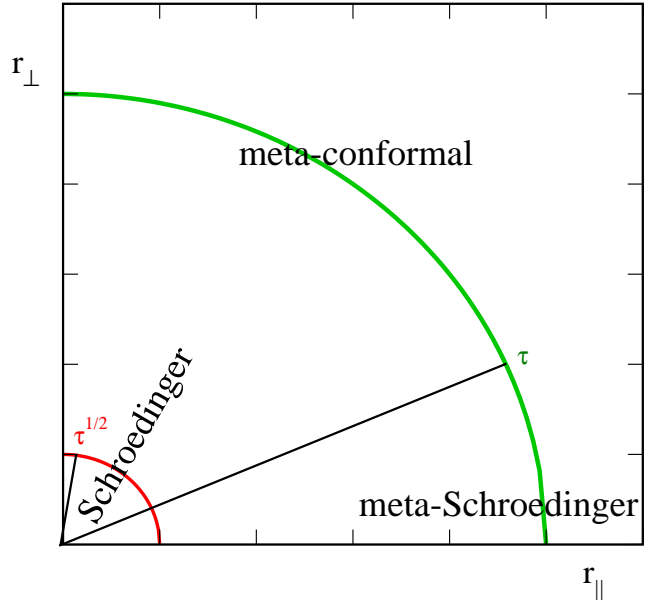}
\end{center}
\caption[fig1]{\small Scaling regimes identified in the directed spherical model for spatial distances $r_{\|}$ and $r_{\perp}$ as compared to
the characteristic length scales $\tau^{1/z}$. 
Conveniently long-ranged initial correlations in the preferred direction (see figure~\ref{fig2}) are necessary for the
existence of the meta-conformal scaling regime. \label{fig1}}
\end{figure}

We have been asking if new scaling regimes can exist in the relaxational dynamics of spin models with a directional bias in the interactions.
The spherical model was used as a non-trivial example in order to better appreciate the conditions under which such new scaling regimes may be
established. Since in this model the dynamic exponent $z=2$, both below and at criticality, the relevant dynamical symmetry is Schr\"odinger-invariance,
which is realised in the scaling regime where $t\to \infty$, $\bigl| \vec{r}\bigr|\to \infty$ isotropically and where $\bigl| \vec{r}\bigr|^2/t$ is kept fixed.
As indicated in table~\ref{tab1}, in the presence of a directional bias either a meta-conformal or else a meta-Schr\"odinger Lie algebra are candidates
for a new dynamical symmetry, which however can only be realised in one of the scaling limits (\ref{gl:metaC}) or (\ref{gl:metaSchr}), respectively.
Figure~\ref{fig1} schematically illustrates the regimes of temporal-spatial distances when either of the new symmetries will be realised instead of
Schr\"odinger-invariance.

What are the supplementary conditions required for either meta-conformal or meta-Schr\"o\-din\-ger-invariance~? We have studied the influence of long-ranged
correlations in the preferred direction, of the form $C\bigl(0; r_{\|},\vec{r}_{\perp}\big)\sim \bigl| r_{\|}\bigr|^{-\aleph}$. 
Since this means in Fourier space
that $\wht{C}(0,q_1,\vec{q}_{\perp})\sim |q_1|^{\alpha}$ with $\alpha=\aleph-1$, 
long-range initial decay is obtained in the region $0<\aleph<1$ (or equivalently
$-1<\alpha<0$). The conditions, for either of these symmetries, are as follows.
\begin{enumerate}
\item \underline{Meta-Schr\"odinger-invariance} is established in the highly anisotropic scaling limit (\ref{gl:metaSchr}), 
see also figure~\ref{fig1}. It predicts the form of
{\em response functions}, which in the Janssen-de Dominicis formalism \cite{Taeu14} can be written as correlators between order-parameter scaling operators $\Phi$ and
response operators $\wit{\Phi}$, for example for the two-point function $R=\bigl\langle \Phi \wit{\Phi}\bigr\rangle$. In the directed spherical model,
we have seen in section~\ref{sec:comparaison} that for quenches to any temperature $T\leq T_c(d)$,
the two-time response can be interpreted accordingly, with the scaling operators $\Phi$ and
$\wit{\Phi}$ being characterised by
\BEQ
\delta=\wit{\delta}=\frac{d+1}{4} \;\; , \;\; \frac{\gamma}{\beta} = \frac{\wit{\gamma}}{\beta} = \demi \;\; , \;\; \xi = - \wit{\xi} = -\frac{\digamma}{2}
\EEQ
without any conditions on the initial correlations. Finally, $\digamma$ is given in table~\ref{tabA2}.
Since $\xi=-\wit{\xi}\ne 0$, the ageing representation of meta-Schr\"odinger-invariance is needed.
\item \underline{Meta-conformal invariance} arises in the isotropic ballistic scaling limit (\ref{gl:metaC}), see again figure~\ref{fig1}. It predicts the form
of {\em correlation functions} of quasi-primary order-parameter scaling operators $\Phi$, 
for example for the two-point function $C=\bigl\langle \Phi\Phi\bigr\rangle$.
In the directed spherical model, this symmetry arises 
\begin{enumerate}
\item at $T<T_c(d)$, for any kind of initial correlations, be they short-ranged or long-ranged. The order-parameter scaling operator $\Phi$ 
is characterised by
\BEQ
\delta = \frac{1+\aleph}{4} \;\; , \;\; \frac{\gamma}{\beta} = \frac{\aleph}{2} \;\; , \;\; \xi = \frac{d-1+\aleph}{4}
\EEQ
\item at $T=T_c(d)$, provided that the initial correlations are relevant. This occurs in the cases (recall $0<\aleph<1$), see also figure~\ref{fig2}
\begin{subequations} \label{gl:SMTc}
\begin{align}
\mbox{\rm\small if~~} 2<d<3-\aleph          &~:~~~ \delta=\frac{\aleph}{2} = \frac{\gamma}{\beta} \;\; , \;\; \xi=\frac{1+\aleph}{4} \\
\mbox{\rm\small if~~} 3-\aleph< d<4-2\aleph &~:~~~ \delta=\frac{\aleph}{2} = \frac{\gamma}{\beta} \;\; , \;\; \xi=1 - \frac{d}{4} \label{gl:SMTc-b}
\end{align}
\end{subequations}
There is no meta-conformal invariance in the directed spherical model if these conditions on the initial correlations are not met.\footnote{Since $\xi\ne 0$, the 
ageing representation of meta-conformal invariance is required.}
\end{enumerate}
This last case (\ref{gl:SMTc}) has an analogue in the other exactly solvable directed model: 
the directed $1D$ Glauber-Ising model, quenched to its critical point $T=0$ \cite{Henkel19a}.
Indeed, for long-ranged initial correlations with $0<\aleph<\demi$, meta-conformal invariance holds for the two-time correlator and
\BEQ \label{gl:GITc}
\delta=\frac{\aleph}{2} = \frac{\gamma}{\beta} \;\; , \;\; \xi=0
\EEQ
We point out the similarity of the behaviour (\ref{gl:SMTc},\ref{gl:GITc}) 
of the $d>2$ critical spherical and the $1D$ Glauber-Ising models. In these cases, sufficiently long-ranged
initial correlations are necessary for the existence of meta-conformal invariance.
\end{enumerate}

A schematic view of these new regimes of dynamical scaling is shown in figure~\ref{fig1} and illustrates the scaling limits
to be taken in either the preferred spatial direction ($r_{\|}$) or else the
transverse spatial direction ($r_{\perp}$) as compared to the characteristic times scales $\tau$, see eqs.~(\ref{gl:metaC},\ref{gl:metaSchr}).
If both $r_{\|}\sim r_{\perp}\sim \tau^{1/2}$ are relatively small, 
then the relaxation dynamics of the spherical model is described by Schr\"odinger-invariance.
However, if larger distances are considered such that anisotropically $r_{\|}\sim \tau$ but $r_{\perp}\sim \tau^{1/2}$, 
and if there is a directional bias in the dynamics, the dynamical symmetry becomes meta-Schr\"odinger-invariance.
Finally, if both distances become isotropically large $r_{\|}\sim r_{\perp}\sim \tau$, 
together with a directional bias and sufficiently long-ranged initial
correlations in the preferred direction, then the dynamical symmetry is meta-conformal invariance.

This should give indications where to look for these new scaling regimes in further spin models,
notably in the two-dimensional directed Ising model \cite{Godreche15b,Godreche18},
and beyond those relatively rare cases where exactly solutions are readily available.
Since the dynamical exponent $z=1$ also arises in the non-equilibrium dynamics of quantum systems \cite{Calabrese07,Calabrese16,Delfino17} 
it would be interesting to
see if in such systems evidence for meta-conformal or meta-Schr\"odinger invariance could be found.

\noindent
{\bf Acknowledgements:} Stoimen Stoimenov tanks the Bulgarian National Science Fund Grant KP-06-N28/6 for financial support.

\newpage

\appsektion{Analysis of the spherical constraint}

The spherical constraint (\ref{eq:SC}) fixes the time-dependence of the function $g(t)$, required for the computation of
correlators and responses, see section~\ref{sec:intro}. The analysis follows and generalises earlier lines \cite{Godr00b,Picone02,Hase06,Ebbi08,Henkel15b}.
In terms of the Laplace-transformation $\lap{g}(p) = \int_0^{\infty} \!\D t\: e^{-pt} g(t)$, the
formal solution is
\BEQ \label{gl:A1}
\lap{g}(p) = \frac{\lap{A}(p)}{1 - 2T \lap{f}(p)}
\EEQ
and depends on the properties of the two auxiliary functions $f(t)$ and $A(t)$ defined in (\ref{eq:fct-aux}).
Tauberian theorems, e.g. \cite{Fell71}, relate the
long-time behaviour of $g(t)$ for $t\to\infty$ to the behaviour of $\lap{g}(p)$ when $p\to 0$.

We begin with the function $f(t)$, defined in eq.~(\ref{eq:fct-aux}). With the dispersion relation (\ref{eq:disper})
for a $d$-dimensional hyper-cubic lattice and ${\cal B}=[-\pi,\pi]^d$
\BEQ
\lap{f}(p) = \frac{1}{(2\pi)^d} \int_{\cal B} \!\D\vec{q} \int_0^{\infty} \!\D t\: e^{-pt} \prod_{a=1}^d e^{-4 (1-\cos q_a) t}
= \int_0^{\infty} \!\D t\: e^{-pt} \left( e^{-4t} I_0(4t)\right)^d
\EEQ
where $I_0$ denotes a modified Bessel function \cite{Abra65}. The small-$p$ expansion of this contains both regular and algebraic terms.
The most simple way to produce them
(slightly heuristically) proceeds as follows. Separate the integration domain into two parts, where $\eta\to\infty$ should be taken at the very end.
Then the leading behaviour for $p$ small is
\BEA
\lap{f}(p) &=& \int_0^{\eta} + \int_{\eta}^{\infty}  \!\D t\; e^{-pt} \left( e^{-4t} I_0(4t)\right)^d \nonumber \\
&\simeq& \int_{0}^{\eta} \!\D t\: \bigl( 1 - p t + \ldots\bigr)  \left( e^{-4t} I_0(4t)\right)^d
+ \int_{\eta}^{\infty} \!\D t\: e^{-pt} \left( \bigl(8\pi t\bigr)^{-1/2}\right)^d
\nonumber \\
&=& A_0 - p A_1 + \ldots + \bigl(8\pi\bigr)^{-d/2} p^{d/2-1} \int_{p\eta}^{\infty} \!\D u\: e^{-u}\, u^{(1-d/2)-1} \nonumber \\
&\simeq& A_0 - p A_1 + \ldots + \bigl(8\pi\bigr)^{-d/2} \Gamma\left( 1 - \frac{d}{2}\right) p^{d/2-1}
\label{gl:A3}
\EEA
where $\Gamma$ is the (analytically continued) Euler Gamma-function \cite{Abra65}.
Herein, in the second line, a small-$p$ expansion was carried out in the first integral, where the constants
$A_i := \int_0^{\infty} \!\D t\; t^{i}  \left( e^{-4t} I_0(4t)\right)^d$ (with $i=0,1,\ldots$)
are finite for $d>2(1+i)$. In this regular part of the expansion, only those terms are retained whose coefficients $A_i$ are finite (see below).
The second integral gives the algebraic part.
For $\eta$ large enough, the large-$x$ asymptotics of $I_0(x)\simeq\frac{e^x}{\sqrt{2\pi x\,}}$ can be used \cite{Abra65}.
Then, in the third line one takes the $p\to 0$ limit (higher-order terms in $p$ could be obtained by successive partial integrations \cite{Copson65}) in order
to obtain the last line and uses an analytical continuation in $d$.
At the very end of the calculation, finally the limit $\eta\to\infty$ is taken.
Eq.~(\ref{gl:A3}) has been derived, mathematically rigorously, many times in the literature \cite{Barber73,Brezin82,Luck85}.

We now turn to the function $A(t)$.
We are interested in long-ranged initial correlations of the asymptotic form $C_n(0)\sim |n|^{-\aleph}$ for $n\to\infty$, in the preferred direction only.
For definite expressions, we shall use the normalised form \cite{Henkel98}, with $\aleph>0$
\BEQ
C_n(0) = \frac{\Gamma( |n|+(1-\aleph)/2)}{\Gamma( |n|+(1+\aleph)/2)} \frac{\Gamma((1+\aleph)/2)}{\Gamma((1-\aleph)/2)} \stackrel{n\to\infty}{\sim} |n|^{-\aleph}
\;\; , \;\;
C_0(0)=1
\EEQ
Consider its Fourier transform $C_n(0)=\frac{1}{2\pi}\int_{\pi}^{\pi}\!\D q\: e^{\II q n} \wht{C}(0,q)$ such that, writing $\alpha:=\aleph-1$ \cite{Henkel19a}
\BEQ
\wht{C}(0,q) = \frac{\Gamma(1+\alpha/2)^2}{\Gamma(1+\alpha)} \left( 2 \sin \frac{|q|}{2}\right)^{\alpha} \stackrel{q\to 0}{\simeq} c_{\alpha} |q|^{\alpha}
\;\; , \;\;
c_{\alpha} = \frac{\Gamma(1+\alpha/2)^2}{\Gamma(1+\alpha)} = \frac{\sqrt{\pi\,}}{2^{\alpha}}\frac{\Gamma(1+\alpha/2)}{\Gamma(\frac{1}{2} + \frac{\alpha}{2})}
\EEQ
Long-range initial correlations are obtained for $\alpha<0$.
The case $\alpha=0$ of short-ranged initial correlations has already been fully analysed in the literature \cite{Godreche13}.
In order to have an algebraic decay in the preferred direction in direct space, we must have $\aleph >0$, hence $-1<\alpha<0$.

Because of the identity $\bigl(2\sin\frac{q}{2}\bigr)^{\alpha} = 2^{\alpha/2} \bigl( 1 - \cos q\bigr)^{\alpha/2}$ for $q\in[0,\pi]$, we have
\BEA
\lap{A}(p) &=& \frac{1}{(2\pi)^d} \int_{\cal B} \!\D\vec{q} \int_0^{\infty} \!\D t\: e^{-pt} c_{\alpha}
               \left( 2 \sin\frac{|q_1|}{2}\right)^{\alpha} \prod_{a=1}^d e^{-4 (1-\cos q_a) t}
\nonumber \\
&=& \frac{c_{\alpha}}{(2\pi)^d} \int_0^{\infty} \!\D t\: e^{-pt} \left( e^{-4t} I_0(4t)\right)^{d-1} 2
    \int_0^{\pi} \!\D q_1\: \left( 2 \sin\frac{q_1}{2}\right)^{\alpha} e^{-4 (1-\cos q_1) t}
\nonumber \\
&=& \frac{c_{\alpha} 2^{1+\alpha/2}}{(2\pi)^d} \int_0^{\infty} \!\D t\: e^{-pt} \left( e^{-4t} I_0(4t)\right)^{d-1}
    \int_0^{\pi} \!\D q_1\:\bigl( 1 - \cos q_1\bigr)^{\alpha/2} e^{-4 (1-\cos q_1) t}
\nonumber \\
&=& \int_0^{\infty} \!\D t\: e^{-pt} \left( e^{-4t} I_0(4t)\right)^{d-1}  {}_1F_1\left( \frac{1+\alpha}{2}, \frac{2+\alpha}{2}; -8t\right)
\EEA
where the identity \cite[(13.2.2)]{Abra65} was used and ${}_1F_1$ is a confluent hypergeometric function \cite{Abra65}.
A spatially decaying long-range initial correlator in the preferred
direction is obtained for $-1<\alpha<0$. The leading small-$p$ behaviour is now found,
in analogy with the above discussion leading to (\ref{gl:A3}), by splitting the integration domain
\BEA
\lap{A}(p) &=& \int_0^{\eta}  +\int_{\eta}^{\infty} \!\D t\: e^{-pt} \left( e^{-4t} I_0(4t)\right)^{d-1}
               {}_1F_1\left( \frac{1+\alpha}{2}, \frac{2+\alpha}{2}; -8t\right)
\nonumber \\
&\simeq& \int_0^{\eta} \bigl( 1 - pt +\ldots\bigr) \left( e^{-4t} I_0(4t)\right)^{d-1}
    {}_1F_1\left( \frac{1+\alpha}{2}, \frac{2+\alpha}{2}; -8t\right) \nonumber \\
& & + \int_{\eta}^{\infty} \!\D t\: e^{-pt}  \bigl( 8\pi t\bigr)^{-(d-1)/2} \frac{\Gamma((2+\alpha)/2)}{\sqrt{\pi\,}} \bigl( 8t\bigr)^{-(1+\alpha)/2}
\nonumber \\
&=& \mathscr{A}_0 - p \mathscr{A}_1 + \ldots + \pi^{\alpha/2} \Gamma\left(\frac{2+\alpha}{2}\right)
\bigl( 8\pi\bigr)^{-(d+\alpha)/2} \int_{p\eta}^{\infty} \!\D u\: e^{-u}\,
\left(\frac{p}{u}\right)^{(d+\alpha)/2} \frac{1}{p}
\nonumber \\
&\simeq& \mathscr{A}_0 - p \mathscr{A}_1 + \ldots + \pi^{\alpha/2} \bigl( 8\pi\bigr)^{-(d+\alpha)/2} \Gamma\left(\frac{2+\alpha}{2}\right)
\Gamma\left( 1 - \frac{d+\alpha}{2}\right) p^{(d+\alpha)/2-1} ~~~
\label{gl:A7}
\EEA
with the constants $\mathscr{A}_i := \int_{0}^{\infty} \!\D t\: e^{-pt} t^i \left( e^{-4t} I_0(4t)\right)^{d-1}
{}_1F_1\left( \frac{1+\alpha}{2}, \frac{2+\alpha}{2}; -8t\right)$
(with $i=0,1,\ldots$) which are finite for $d+\alpha>2(1+i)$. In the regular part of the expansion,
only those terms are retained whose coefficients $\mathscr{A}_i$ are finite.
For the asymptotics of ${}_1F_1(-x)$ for $x\to +\infty$ in the second integral of the second line,
the identity \cite[(13.5.1)]{Abra65} was used.  On the explicit forms in the third line,
perform the $p\to 0$ limit.
At the very end of the calculation, the limit $\eta\to\infty$ is taken.

\begin{table}
\begin{center}\begin{tabular}{|l|ll|} \hline
~$d$~          & \multicolumn{1}{|c}{$~\lap{f}(p)$~} & \multicolumn{1}{c|}{~$\lap{A}(p)$~} \\[0.05truecm] \hline
$2<d<2-\alpha$   & $A_0 + \bigl(8\pi\bigr)^{-1/2} \Gamma\bigl(1-d/2\bigr) p^{d/2-1}$
               & $\bigl(8\pi\bigr)^{-(d+\alpha)/2} \pi^{\alpha/2} \Gamma\bigl(\frac{2+\alpha}{2}\bigr)
                 \Gamma\bigl(1-\frac{d+\alpha}{2}\bigr) p^{(d+\alpha)/2-1}$ \\[0.05truecm]
$2-\alpha<d<4$ & $A_0 + \bigl(8\pi\bigr)^{-1/2} \Gamma\bigl(1-d/2\bigr) p^{d/2-1}$
               & $\mathscr{A}_0 + \bigl(8\pi\bigr)^{-(d+\alpha)/2} \pi^{\alpha/2}
                 \Gamma\bigl(\frac{2+\alpha}{2}\bigr)\Gamma\bigl(1-\frac{d+\alpha}{2}\bigr) p^{(d+\alpha)/2-1}$ \\[0.05truecm]
$4<d<4-\alpha$ & $A_0 - A_1 p + \ldots$
               & $\mathscr{A}_0 + \bigl(8\pi\bigr)^{-(d+\alpha)/2} \pi^{\alpha/2} \Gamma\bigl(\frac{2+\alpha}{2}\bigr)
                 \Gamma\bigl(1-\frac{d+\alpha}{2}\bigr) p^{(d+\alpha)/2-1}$ \\[0.05truecm]
$4-\alpha<d$   & $A_0 - A_1 p + \ldots$
               & $\mathscr{A}_0 - \mathscr{A}_1 p + \ldots$ \\[0.05truecm] \hline
\end{tabular}
\caption[tabA1]{Leading small-$p$ behaviour of the auxiliary functions $\lap{f}(p)$ and $\lap{A}(p)$, for dimensions $d>2$ and $-1<\alpha<0$. \label{tabA1}}
\end{center}
\end{table}

In summary, the small-$p$ behaviour of the two auxiliary functions is listed in table~\ref{tabA1}.

Returning to (\ref{gl:A1}), two cases must be distinguished.
\begin{enumerate}
\item \underline{$\lap{g}(p)$ has a p\^ole at some $p_0>0$}. Since $\lap{A}(p)$ is finite for all $p>0$,
then $p_0$ is given by $1-2T\lap{f}(p_0)=0$ or else
\BEQ
\lap{f}(p_0) = \frac{1}{2T}
\EEQ
Around $p_0$, the behaviour of $\lap{g}(p)$ is
\BEQ
\lap{g}(p) \simeq - \frac{\lap{A}(p_0)}{2T \lap{f}^{\:'}(p_0)}\frac{1}{p-p_0} ~~\Longrightarrow~~
g(t) \sim g_0 e^{t/\tau_{\rm r}} \mbox{\rm ~~with $\frac{1}{\tau_{\rm r}}=p_0$}
\EEQ
This is the characteristic exponential behaviour for temperatures $T>T_c(d)>0$, with a finite relaxation time $\tau_{\rm r}$ \cite{Godreche13} which does not
depend on the initial correlations. For $d>2$,
the critical temperature $T_c(d)$ is found from the $p_0\to 0$ limit
\BEQ
\frac{1}{2T_c(d)} = \lap{f}(0) = \int_0^{\infty} \!\D t\: e^{-pt} \left( e^{-4t} I_0(4t)\right)^d
\EEQ
which is easily computed numerically. It is known that
$\frac{1}{T_c(3)}=\frac{\sqrt{3\,}\,-1}{192 \pi^3}\Gamma^2\bigl(\frac{1}{24}\bigr)\Gamma^2\bigl(\frac{11}{24}\bigr)$
\cite{Cara03}, \cite[eq. (6.612.6)]{Grad07}.
\item \underline{$\lap{g}(p)$ has a singularity at $p=0$}. One may insert the expansions from table~\ref{tabA1} into (\ref{gl:A1}) to obtain
\BEQ \label{gl:A11}
\lap{g}(p)\simeq \frac{\mathscr{A}_0 - \mathscr{A}_1 p + a_{\alpha} p^{(d+\alpha)/2-1}}{1 - T/T_c + 2T A_1 p - 2T f_d p^{d/2-1}}
\EEQ
and where $a_{\alpha}$ and $f_d$ denote the amplitudes of the algebraic parts of $\lap{A}(p)$ and $\lap{f}(p)$,
respectively. The total magnetisation is $M^2 = 1 -T/T_c$.

First, at the critical point $T=T_c$, the magnetisation $M=0$. This produces, to leading order, $\lap{g}(p)$
and Laplace-inversion leads to the sought long-time behaviour of $g(t)$ \cite{Picone02}
\BEQ
\lap{g}(p) \simeq \left\{
\begin{array}{l} - \frac{a_{\alpha}}{2T_c f_d} p^{\alpha/2} \\[0.09truecm]
-\frac{\mathscr{A}_0}{2T_c f_d} p^{1-d/2} \\[0.09truecm]
\frac{\mathscr{A}_0}{2T_c A_1} p^{-1}
\end{array} \right.
~~\Longrightarrow~~
g(t) \simeq \left\{
\begin{array}{ll} - \frac{a_{\alpha}}{2T_c f_d} \frac{1}{\Gamma\bigl(1+\frac{\alpha}{2}\bigr)} t^{-1-\alpha/2} & \mbox{\rm if $2<d<2-\alpha$} \\
-\frac{\mathscr{A}_0}{2T_c f_d} \frac{1}{\Gamma\bigl(2-\frac{d}{2}\bigr)} t^{d/2-2}                            & \mbox{\rm if $2-\alpha<d<4$} \\
\frac{\mathscr{A}_0}{2T_c A_1} t^0                                                                             & \mbox{\rm if $4<d$}
\end{array} \right.
\EEQ
Second, below the critical point, one has $T<T_c$, hence $M^2=1-T/T_c>0$. From (\ref{gl:A11}), it follows that to lowest non-trivial order
\BEA
\lap{g}(p) &\simeq& \frac{1}{M^2} \biggl( \mathscr{A}_0 - \mathscr{A}_1 p + a_{\alpha} p^{(d+\alpha)/2-1} \biggr)
\left( 1 - \frac{2T A_1}{M^2} p + \frac{2Tf_d}{M^2} p^{d/2-1} \right)  \\
&\simeq& \frac{1}{M^2} \left\{
\begin{array}{ll} a_{\alpha} p^{(d+\alpha)/2-1} & \mbox{\rm if $2<d<2-\alpha$} \\[0.07truecm]
\mathscr{A}_0 + a_{\alpha} p^{(d+\alpha)/2-1}   & \mbox{\rm if $2-\alpha<d<4-\alpha$} \\[0.07truecm]
\mathscr{A}_0 - \biggl( \mathscr{A}_1 + 2 T \mathscr{A}_0 A_1/M^2\biggr) p + a_{\alpha} p^{(d+\alpha)/2-1}   & \mbox{\rm if $4-\alpha<d$}
\end{array} \right. \nonumber
\EEA
with the corresponding long-time behaviour
(the singular term $\sim\mathscr{A}_0$ is only present for $d>2-\alpha$ and we can drop an eventual singular term
$\sim p$ (which will lead to a term $\sim \delta'(t)$ in $g(t)$) since it merely produces corrections to scaling in the correlator)
\BEQ \label{gl:A.14}
g(t) \simeq \frac{\mathscr{A}_0}{M^2}\delta(t) + \frac{a_{\alpha}}{M^2} \frac{1}{\Gamma\bigl( 1 +\frac{d+\alpha}{2}\bigr)} t^{-(d+\alpha)/2}
\EEQ
\end{enumerate}

\begin{table}
\begin{center}\begin{tabular}{|lr|l|} \hline
\multicolumn{2}{|c}{conditions} & \multicolumn{1}{|c|}{~$\digamma$~} \\ \hline
$T=T_c$ & $2<d<2-\alpha$        & $-1 - \alpha/2$ \\
$T=T_c$ & $2-\alpha < d < 4$    & ~~~$d/2-2$ \\
$T=T_c$ & $4 < d$               & ~~~$0$ \\[0.2cm]
$T<T_c$ & $2<d$                 & $-(d+\alpha)/2$ \\ \hline
\end{tabular}
\caption[tabA2]{Exponent $\digamma$ defined in (\ref{gl:A15}) for temperatures $T\leq T_c$.\label{tabA2}}
\end{center}
\end{table}

It follows that for $T\leq T_c$, the long-time behaviour is always algebraic
\BEQ \label{gl:A15}
g(t) \simeq g_{\infty}  t^{\digamma}
\EEQ
The values of the exponent $\digamma$ are listed in table~\ref{tabA2}.\footnote{The kinetic spherical model with spatially isotropic long-range
initial correlations was analysed long ago \cite{Picone02}. Here, long-range initial conditions are only assumed in the single preferred direction. Hence the values of
$\digamma$ in table~\ref{tabA2} are different from those in \cite{Picone02}.}
Its is always non-positive but may depend via $\alpha$ on the
initial condition. The (non-universal) amplitude $g_{\infty}$ in (\ref{gl:A15}) almost always cancels when evaluating correlations or responses.
At $T=T_c$, the exponent $\digamma$ appears to change continuously between the regimes,
but precisely at the border-line dimensions additional logarithmic
factors may occur in $g(t)$ which we did not work out explicitly.


\end{document}